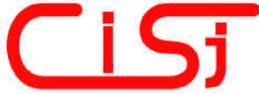

# BRAIN TISSUES SEGMENTATION ON MR PERFUSION IMAGES USING CUSUM FILTER FOR BOUNDARY PIXELS

**Svitlana Alkhimova [1), Andrii Krenevych [2)**

[1) National Technical University of Ukraine "Igor Sikorsky Kyiv Polytechnic Institute",
37, Prosp.Peremohy, Solomyanskyi district, Kyiv, Ukraine, 03056, asnarta@gmail.com
[2) Taras Shevchenko National University of Kyiv
64/13, Volodymyrska Street, Kyiv, Ukraine, 01601, krenevych@knu.ua

**Abstract:** The fully automated and relatively accurate method of brain tissues segmentation on T2-weighted magnetic resonance perfusion images is proposed. Segmentation with this method provides a possibility to obtain perfusion region of interest in images with abnormal brain anatomy that is very important for perfusion analysis. In the proposed method the result is presented as a binary mask, which marks two regions: brain tissues pixels with unity values and skull, extracranial soft tissue and background pixels with zero values. The binary mask is produced based on the location of boundary between two studied regions. Each boundary point is detected with CUSUM filter as a change point for iteratively accumulated points at time of moving on a sinusoidal-like path along the boundary from one region to another. The evaluation results for 20 clinical cases showed that proposed segmentation method could significantly reduce the time and efforts required to obtain desirable results for perfusion region of interest detection on T2-weighted magnetic resonance perfusion images with abnormal brain anatomy.

## 1. INTRODUCTION

Dynamic susceptibility contrast (DSC) perfusion magnetic resonance (MR) imaging has already been widely used for the management of patients with brain tumors and cerebrovascular diseases, such as vascular stenosis or stroke [1-3]. It is based on the fact that T2-weighted MR images show decreased signal intensity while the iodinated contrast agent passes through the tissue. DSC exam output is a series of T2 weighted images acquired before, during, and after iodinated contrast agent injection into the vascular system. This time-sequence data reflect local changes in perfusion tissue characteristics. To estimate perfusion tissue characteristics according to the obtained time series it is necessary to carry out analysis of signal intensity changes on a pixel-by-pixel basis within the same part of the human body.

The results of DSC perfusion exam can be used both for quantitative estimation of perfusion characteristics as well as visual assessment of tissue perfusion. It is visualization of perfusion characteristics on the so-called perfusion maps that serves to detect regions with potential perfusion lesions and determine a diagnosis. However, brain lesions can be poorly visualized on perfusion maps due to the low contrast between the lesion and surrounding tissues. Such effect occurs when perfusion maps display extremely high values for noised pixels or skull and extracranial soft tissues regions. Thus, the important step of DSC perfusion analysis is pre-processing of time-sequence data through brain tissues segmentation and binary mask creation for so-called perfusion region of interest (ROI) [4-7].

## 2. PROBLEM STATEMENT

Present software applied for medical image processing uses different methods of brain tissues segmentation. There is manual and more or less automated segmentation, but the last one is more preferable for clinical use [4]. Despite the fact that manual segmentation gives more accurate results it is more laborious and time-consuming. It should be mentioned that to obtain accurate segmentation results an operator has to have sufficient knowledge and experience to identify brain anatomical structures and its lesions on MR images. However, regarding the disadvantages of manual segmentation mentioned here, the automated segmentation is not that ideal for this task [6, 7]. In order to be of practical use, the segmentation method should provide accurate results for images with abnormal brain anatomy. It is automated segmentation methods that generally give wrong perfusion ROI for such images, as a result, segmentation faults lead to the absence of perfusion characteristics data for pixels in regions with potential perfusion lesions. Therefore, there are a number of different methods of brain tissues segmentation on MR images [8-12].

Most of the automated brain tissues segmentation methods are used for T1-weighted MR images processing. For clarity, these methods can be divided into two groups: those which are based on pixels intensity analysis (thresholding and clustering algorithms), and those which use patterns (neural network classifiers and atlas-based algorithms). The central problem of the first group methods is overlapping pixel intensity values in region of interest and background. The second group problem is lack of age-sex-race-specific pre-segmented template data and lack of training samples for different size, density, and volume of brain lesions.

Surrounding brain tissues appear as bright pixels on T2-weighted MR perfusion images. It is the main reason for segmentation faults while using automated methods which work fine with T1-weighted images [13]. The proposed strategy of parameter-based transformation of T2-weighted into T1-weighted image intensities [14] solves partially the issue of using suitable for T1-weighted images methods to segment T2-weighted images. However, it does not solve the problem of segmentation faults for images with abnormal brain anatomy.

The idea of using CUSUM filter to track the boundary for autonomous vehicles [15, 16] was applied to solve image segmentation issue in general [17]. The main purpose of CUSUM filter usage is to solve the issue of boundary tracking at time of movement between two regions using only local information in the presence of noise. Thus, it gives the opportunity to solve the segmentation issues in different domains [18, 19]. But in order to get accurate segmentation results using CUSUM filter in specific domain relationship between two disparate regions should be established and appropriate movements pattern should be developed so that the boundary can be accurately tracked and estimated [17, 18].

## 3. THE AIM AND OBJECTIVES OF THE STUDY

The aim of the current study is to develop a fully automated method of brain tissues segmentation on T2-weighted MR perfusion images of a human head with abnormal brain anatomy using CUSUM filter for boundary pixels.

To achieve the aim, the following tasks have been set:

− to develop an algorithm for path planning with specific movements pattern along the boundary between two regions: one region is specified with brain tissues pixels and the other one with skull, extracranial soft tissues and background pixels;

− to adapt CUSUM filter for a change point detection in order to determine the moment of boundary moving from one region to another;

− to evaluate the results of brain tissues segmentation on T2-weighted MR perfusion images of a human head with abnormal brain anatomy using CUSUM filter for boundary pixels.

## 4. SEGMENTATION WITH CUSUM FILTER FOR BOUNDARY PIXELS

The proposed segmentation method of T2-weighted MR perfusion images produces a binary mask image which marks brain tissues.

After sorting analysed images as 4-dimensional time-series segmentation process is performed using the $4^{th}$ image for every spatial position (T2-weighted images acquisition protocol can vary, so processed image should be selected after discarding the first few time-points images at which signal intensity is not reached a steady state).

The binary mask, which marks brain tissues, is produced based on the location of boundary B. Boundary B separates two image regions: region $\Omega_1$ is specified with brain tissues pixels and region $\Omega_2$ is specified with skull, extracranial soft tissues and background pixels. Thus, image is represented by $\Omega$, so that $\Omega = \Omega_1 \cup \Omega_2 \cup B$ and $\Omega_1 \cap \Omega_2 = \emptyset$.

In the proposed method the boundary B tracking is provided by detecting image points close to the boundary of the studied regions. Each mentioned image points is detected with CUSUM filter as a change point for iteratively accumulated points at time of moving on a path along the boundary from one region to another. The motion trajectory is a sinusoidal-like path which is defined by iterative change of movement direction in region $\Omega_1$ in order to reach region $\Omega_2$, and vice versa.

### 4.1 INITIAL POINT DETECTION

To start moving on a sinusoidal-like path along the boundary between two regions, the initial starting point $w_0(x_0, y_0)$ should be detected, which definitely belongs to boundary B. The simplest way to detect such point is to do it manually.

To provide an automatic approach of segmentation the initial point $w_0(x_0, y_0)$ detection is done iteratively through analysing diagonal pixels in turns. The analysis starts with the angle which corresponds to the left posterior part of the visualization object and continues along the image diagonal towards its centre.

Mean value of pixels intensity from a local neighbourhood of analysed pixel is compared with set threshold. The neighborhood is defined in the form of a 3x3 mask with the analysed pixel in the center. If the mean value of pixels intensity is greater than threshold, $w_0$ point is considered to be detected (that corresponds to the analyzed pixel).

For $w_0$ point detection process threshold value is specified as intensity value obtained with Otsu method, so that set threshold maximizes the inter class variance of two classes of image pixels (foreground pixels and background pixels).

## 4.2 PATH PLANNING WITH SPECIFIC MOVEMENTS PATTERN

After $w_0$ point is detected the movement on a path starts from this point along the boundary between $\Omega_1$ and $\Omega_2$ regions in the counterclockwise direction.

Every new point $w_k(x_k, y_k)$ on a path along the boundary between two regions is detected using:

$$w_k = w_{k-1} + V \cdot n°, \qquad (1)$$

where V – distance between two neighboring image points on motion trajectory, $n°$ – unit vector that makes an angle $\theta_{k-1}$ with the positive x-axis.

The angle $\theta$ is calculated as follows:

$$\theta_k = \theta_{k-1} + \theta_k^{step} + \theta_k^{correction}, \qquad (2)$$

where $\theta_k^{step}$ – angle change value at each step k of motion trajectory creation, $\theta_k^{correction}$ – angle correction value, that is necessary to avoid an issue with infinite looping inside one region.

The value of angle $\theta_k^{step}$ is calculated as follows:

$$\theta_k^{step} = \begin{cases} (-1)^{I_{\Omega_2}(w_{k-1})} \cdot \dfrac{\pi}{4}, & \text{for } w_{k-1} \notin B; \\ (-1)^{I_{\Omega_1}(w_{k-1})} \cdot \dfrac{\pi}{2}, & \text{for } w_{k-1} \in B, \end{cases} \qquad (3)$$

where $I_\Omega$ – the indicator function of region $\Omega$. Angle change values for the proposed method were selected as a rule of thumb based on the analysis of motion trajectory data obtained on T2-weighted MR perfusion images.

To detect the maximum number of points, which belong to boundary B, and at the same time minimize the number of analysed pixels, in the Eq. (3) double angle is used for changing the direction of movement when the boundary moves from $\Omega_1$ to $\Omega_2$, and vice versa (Fig.1).

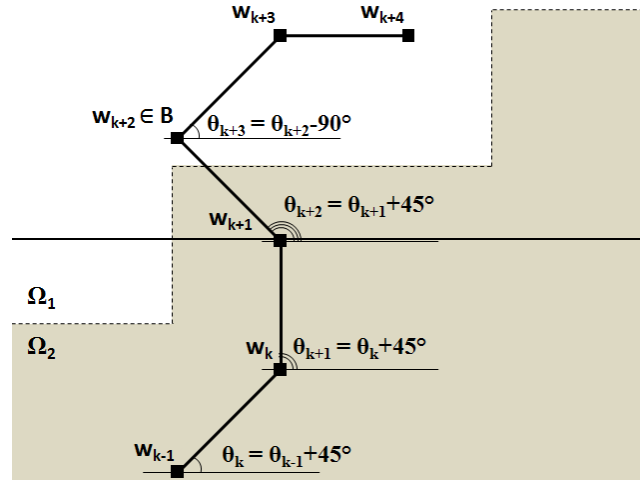

**Figure 1 – Basic principle of double angle usage for changing the direction of movement when the boundary moves from one region to another**

In order to avoid infinite looping inside one region, e.g., motion trajectory makes a full circle inside one region without returning to the boundary (either inside region $\Omega_1$ or inside region $\Omega_2$), correction value $\theta_k^{correction}$ is used in the Eq. (2). It is calculated by the following equation:

$$\theta_k^{correction} = \theta_k^{shift} \cdot I_{\{w_k = w_{k-9}\}}, \qquad (4)$$

where $I_{\{w_k = w_{k-9}\}}$ – indicator that the motion trajectory makes a full circle, $\theta_k^{shift}$ – angle shift value, that is calculated by the following equation:

$$\theta_k^{shift} = (-1)^{I_{\Omega_1}(w_k)} \cdot \dfrac{\pi}{3}. \qquad (5)$$

In the current study, the value of distance V between two neighboring image points is a constant value, and it is set equal to 0.39 minimal side length of the image pixel spacing.

Motion with subpixel accuracy allows in most cases to detect points which belong to boundary B, as well as those located in neighboring pixels of the image. This, in its turn, allows to have connected neighbors for pixels that belong to the boundary between two regions and therefore simplifies further binary masking creation for brain tissues. In addition, this approach reduces error probability because of losing boundary pixels.

The minimum acceptable value of distance V, in order to definitely get outside the one-pixel region, can be calculated on the basis of a geometric shape of motion trajectory. It is approximately equal to 0.3827 minimal side length of the image pixel spacing. The further reducing of V value will increase cases of infinite looping inside one region (inside one pixel of the same region in the current case), and therefore, it will take more processing time for motion trajectory creation due to the involving of $\theta_k^{correction}$ value in calculations.

All of the aforesaid describes the iterative algorithm of a motion trajectory creation as a sinusoidal-like path between two studied regions.

## 4.3 CHANGE POINT DETECTION WITH CUSUM FILTER

Differentiation of image points on motion trajectory as points which belong to boundary B or as points inside region $\Omega_1$ or $\Omega_2$ is provided with CUSUM filter.

CUSUM filter uses cumulative sums to detect the moment of time when statistical characteristics of the monitored process undergo a change, i.e., to detect change point. According to the original CUSUM rule proposed by Page in [20] a change point detection is provided as follows:

$$\tau = \min(k > 0 : S_k > h), \qquad (6)$$

where $\tau$ – alarm time, that is, the time when a change is detected and process is out of control, h – given threshold controlling the false alarm rate that should be chosen to minimize both false-alarms and time to change point detection, $S_k$ – cumulative sum up to and including kth statistical characteristics value of monitored process.

The cumulative sum value $S_k$ is based on the log-likelihood ratio and defined as follows:

$$S_k = \max\left(0, S_{k-1} + \ln\left[\frac{g_1(z_k)}{g_0(z_k)}\right]\right), \qquad (7)$$

where $g_0$ is the probability density function of the monitored process characteristics $z_k$ before the change point and $g_1$ the density after the change point. Thus, on each step the value of cumulative sum is compared with the threshold h, and if the threshold is exceeded, the signal about change point is initiated.

In the proposed method a sequence of image intensities at points on a path along the boundary between two regions $z_0 = I(w_0)$, $z_1 = I(w_1)$, …, $z_k = I(w_k)$ are used as statistical characteristics of the monitored process. Here each value is obtained with bilinear interpolation of original image pixel data at given spatial location on a path.

In case of image processing with CUSUM filter, the actual data distribution of monitored process is unknown, as a result, probability density functions $g_0$ and $g_1$ are unknown as well. For this reason, in the proposed method the log-likelihood ratio for cumulative sum calculation replaced by a score function G, which is sensitive to expected changes. The score function G is computed for $\Omega_1$ and $\Omega_2$ regions as follows:

$$G_j(w_k) = I(w_k) - \mu_j, \quad j = 1,2, \qquad (8)$$

where $I(w_k)$ – intensity value at point $w_k$, $\mu_j$ – mean intensity value for sequence of q points prior to point $w_k$ in $\Omega_j$ region. In the current study, the value of q is set equal to 45.

To detect the change point the proposed method uses CUSUM filter with cumulative sum calculation as follows:

$$\begin{cases} S_{\tau_i} = 0; \\ S_k = \max[0, S_{k-1} - I_{\Omega_1} G_1(w_k) + I_{\Omega_2} G_2(w_k)], \end{cases} \qquad (9)$$

where $k > \tau_i$, $I_\Omega$ – the indicator function of region $\Omega$, G – score function, which is sensitive to expected change, $\tau_i$ – sequence of alarm time moments for a change point detection, which is calculated by the following equation:

$$\begin{cases} \tau_0 = 0; \\ \tau_{i+1} = \min[k > \tau_i : S_k > h], \quad i \geq 0, \end{cases} \qquad (10)$$

where h – threshold controlling the false alarm rate, which is calculated by the following equation:

$$h = |\mu_1 - \mu_2|, \qquad (11)$$

where $\mu_1$ and $\mu_2$ – mean intensity values for sequence of 45 points prior to point $w_k$ in $\Omega_1$ and $\Omega_2$ regions respectively.

In case of processing initial points on a path between two regions, the number of points prior to point $w_k$ is not sufficient to calculate the threshold value with Eq. (11). Threshold value for such points processing is set equal to the value obtained applying Otsu method to the whole image.

According to the proposed Eq. (9) the initial CUSUM filter value starts at zero. If the data after a change point are available cumulating of the sum starts from zero. In order to improve performance of the proposed CUSUM filter, reset of cumulative sum is provided in consonance with "fast initial response" approach [21] and defined as follows:

$$S_{\tau_p} = \frac{1}{2} \cdot S_{\tau_{p-1}-1}, p \geq 2. \qquad (12)$$

The proposed segmentation method stops a process of change point detection in two cases: motion trajectory goes back to the pixels with initial starting point $w_0$ location, or motion trajectory reaches the image border.

The sequence of detected change points determines image pixels that belong to the boundary B between two studied regions. The binary mask which marks brain tissues is produced on the basis of these pixels. Values of binary masks are unity for brain tissues pixels and zero for skull, extracranial soft tissues and background pixels.

## 5. RESULTS AND DISCUSSION

The proposed method of segmentation was applied to detect brain tissues and to create a binary mask for the so-called perfusion ROI on T2-weighted MR perfusion images of a human head with abnormal brain anatomy from 20 clinical cases.

The results shown here are in whole based upon data generated by the TCGA Research Network: http://cancergenome.nih.gov/.

In order to evaluate the proposed method, the results of brain tissues segmentation were compared with a reference standard, which is the manually marked brain perfusion ROI by two experienced radiologists in perfusion MR imaging.

Metrics value of Dice similarity index, sensitivity, specificity and accuracy were calculated to evaluate the segmentation results.

These metrics are based on the fact that each pixel of the segmented image is assigned to one of the following categories:
− false positive (FP) pixels, i.e., those which are segmented as perfusion ROI pixels, but they are not;
− false negative (FN) pixels, i.e., those which are segmented as pixels not of the perfusion ROI, but they are;
− true positive (TP) pixels, i.e., those which are segmented as perfusion ROI pixels, and they are;
− true negative (TN) pixels, i.e., those which are segmented as pixels not of the perfusion ROI, and they are not.

Dice similarity index (also known as Dice coefficient) was used to estimate spatial overlap between two regions (image region, which is identified automatically with brain tissues segmentation using CUSUM filter for boundary pixels and image region, which is identified manually as brain perfusion ROI by two experienced radiologists in consensus) and calculated as follows:

$$D = \frac{2 \cdot TP}{2 \cdot TP + FP + FN}. \qquad (13)$$

The values of this metric can vary from 0 (no overlap) to 1 (perfect agreement).

The sensitivity and specificity were used to estimate the number of properly detected perfusion ROI pixels.

The sensitivity or true positive fraction (TPF) shows the ability of the proposed segmentation method to correctly detect pixels, which belong to the perfusion ROI. It is calculated as follows:

$$TPF = \frac{TP}{TP + FN}. \qquad (14)$$

The sensitivity values can vary from 0 to 1: the higher sensitivity shows the lower missed true pixels of perfusion ROI.

The specificity or true negative fraction (TNF) shows the ability of the proposed segmentation method to correctly detect pixels, which do not belong to the perfusion ROI (skull, extracranial soft tissues and background pixels in the current case). It is calculated as follows:

$$TNF = \frac{TN}{TN + FP}. \qquad (15)$$

The specificity values can vary from 0 to 1: higher specificity shows the lower missed true pixels of skull, extracranial soft tissues and background.

The accuracy shows the rate of the pixels that are correctly identified with the proposed segmentation method. This metric is calculated as follows:

$$ACC = \frac{TP + TN}{TP + FP + TN + FN}. \qquad (16)$$

The values of this metric can vary from 0 (in case of non-agreement) to 1 (in case of complete agreement of obtained results and reference standard).

Evaluation results of the proposed segmentation method of brain tissues segmentation on T2-weighted MR perfusion images of a human head with abnormal brain anatomy are shown in Table 1.

**Table 1. Evaluation results (average value and standard deviation) of brain tissues segmentation using CUSUM filter for boundary pixels**

| Clinical case | Metrics | | | |
|---|---|---|---|---|
| | Dice index (D) | Sensitivity (TPF) | Specificity (TNF) | Accuracy (ACC) |
| 1 | 0.9745±0.0087 | 0.98±0.026 | 0.9912±0.0049 | 0.9892±0.0029 |
| 2 | 0.9494±0.0157 | 0.9273±0.0339 | 0.9936±0.0022 | 0.9803±0.0056 |
| 3 | 0.9784±0.0166 | 0.9736±0.0365 | 0.9947±0.0024 | 0.9895±0.0074 |
| 4 | 0.9891±0.0038 | 0.9961±0.004 | 0.9918±0.0036 | 0.9931±0.0022 |
| 5 | 0.9869±0.0048 | 0.9952±0.0014 | 0.9917±0.0033 | 0.9927±0.0028 |
| 6 | 0.985±0.0021 | 0.9955±0.006 | 0.9894±0.0039 | 0.9911±0.0012 |
| 7 | 0.959±0.0123 | 0.9673±0.0345 | 0.9837±0.0147 | 0.9802±0.0075 |
| 8 | 0.9848±0.0009 | 0.9974±0.0024 | 0.9899±0.0015 | 0.9919±0.0008 |
| 9 | 0.9886±0.0021 | 0.9955±0.0031 | 0.9939±0.0032 | 0.9944±0.0018 |
| 10 | 0.9763±0.0065 | 0.9907±0.0116 | 0.9872±0.0066 | 0.9883±0.0038 |
| 11 | 0.9795±0.0154 | 0.9882±0.0163 | 0.9896±0.0049 | 0.9893±0.0078 |
| 12 | 0.9801±0.0069 | 0.9944±0.0068 | 0.9875±0.0055 | 0.9896±0.0036 |
| 13 | 0.9798±0.0039 | 0.9852±0.0115 | 0.9918±0.0031 | 0.9905±0.001 |
| 14 | 0.9864±0.0032 | 0.992±0.0084 | 0.9918±0.002 | 0.9918±0.002 |
| 15 | 0.9736±0.0183 | 0.9924±0.0081 | 0.9864±0.007 | 0.9878±0.0073 |
| 16 | 0.9619±0.0212 | 0.952±0.0545 | 0.9946±0.0048 | 0.99±0.0031 |
| 17 | 0.9703±0.0136 | 0.9565±0.0246 | 0.9956±0.003 | 0.987±0.0065 |
| 18 | 0.9647±0.0394 | 0.9557±0.0721 | 0.9919±0.0014 | 0.9815±0.0213 |
| 19 | 0.9624±0.0365 | 0.9505±0.0672 | 0.9925±0.0035 | 0.9815±0.0183 |
| 20 | 0.9576±0.0229 | 0.945±0.0469 | 0.9928±0.0025 | 0.9833±0.0083 |

Three sample images were selected for presentation of the segmentation results, i.e., the image with the lowest (Fig. 2), average (Fig. 3), and highest (Fig. 4) similarity.

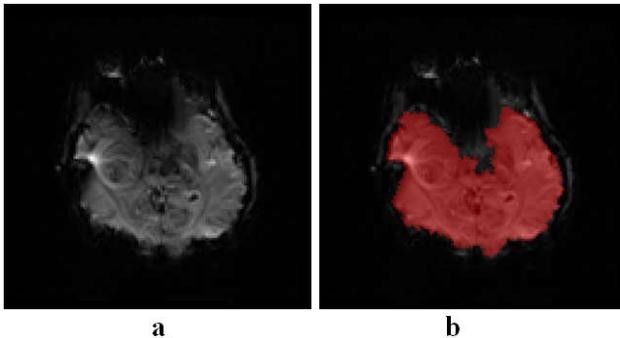

a  b
**Figure 2 – Segmentation results on the selected sample with the lowest similarity to the reference standard: a - original image; b - perfusion ROI extracted from a.**

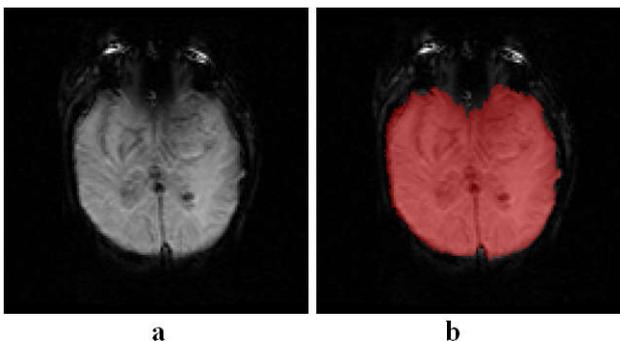

a  b
**Figure 3 – Segmentation results on the selected sample with the average similarity to the reference standard: a - original image; b - perfusion ROI extracted from a.**

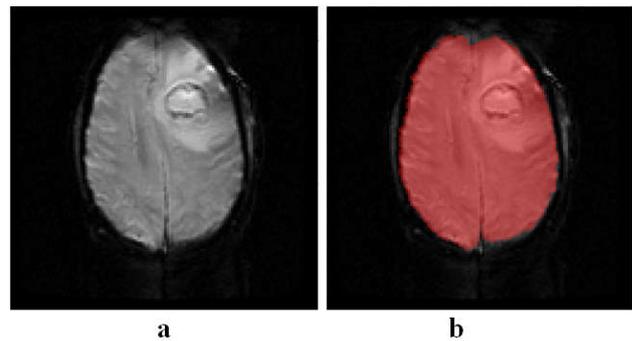

a  b
**Figure 4 – Segmentation results on the selected sample with the highest similarity to the reference standard: a - original image; b - perfusion ROI extracted from a.**

Obtained segmentation results are in good agreement with the reference standard with a Dice index values 0.9744 ± 0.0216, have high values of sensitivity (0.9765 ± 0.0411), specificity (0.9911 ± 0.0052), and accuracy (0.9881 ± 0.0101).

All detected perfusion ROIs were deemed by two experienced radiologists as satisfactory enough for clinical use. This gives the reasons to consider the segmentation results with the use of proposed method as very satisfactory for quantitative measurements and visual assessment of brain perfusion data and does not need to be manually edited much by radiologists.

Finally, obtained in the current study results of brain tissues segmentation were compared with the results obtained using thresholding method. Thresholding is a state-of-the-art method to detect brain tissues on T2-weighted MR perfusion images.

It is widely used in perfusion analysis system for clinical use. The comparison was performed on the same images that were selected for the evaluation experiments. To make comparison more meaningful, threshold value was defined to present segmentation results with the highest similarity to the reference standard. For this purpose, each threshold value was selected via brute force thresholding from the intensities range of each analyzed image. Developed segmentation method showed superior performance in the current study compared with the threshold method that reached only $0.9274 \pm 0.0245$ for evaluating with Dice index metric. Despite the relatively high Dice index values of the thresholding method, it showed low values of sensitivity ($0.8675 \pm 0.0413$). Such results prove the ability of the proposed method to detect perfusion ROI on images with abnormal brain anatomy.

Average processing time of the proposed method is about 0.2s per image of 128x128 pixels in size (Intel Core i5-460M 2.53 GHz, single threaded), which is quite fast. It can play an important role in cases that require processing of a large number of patient images.

As shown in the results, the proposed method can reliably detect the perfusion ROI on T2-weighted MR perfusion images of a human head with abnormal brain anatomy.

## 6. CONCLUSIONS

The current study resulted in the development of a fully automated method of brain tissues segmentation on T2-weighted MR perfusion images of a human head with abnormal brain anatomy using CUSUM filter for boundary pixels. Thus, the major findings of the current study are presented below.

1. The algorithm for path planning with specific movement pattern along the boundary between two regions was developed: one region is specified with brain tissues pixels and the other one with skull, extracranial soft tissues and background pixels. The motion trajectory was developed as sinusoidal-like path which was defined by iterative change of movement direction inside one region in order to reach another region. The problem of maximization of detected points, which belong to the boundary between two regions, was solved and at the same time the number of processed pixels using the algorithm was minimized. The problem of infinite looping inside one region was solved to predict cases of making a full circle by motion trajectory inside one region without returning to the boundary.

2. The CUSUM filter was adapted for a change point detection to determine the moment when boundary moves from one region to another. To calculate the cumulative sum with proposed CUSUM filter the log-likelihood ratio was replaced by a score function, which is sensitive to expected change of intensity values at time of moving on a path along the boundary from one region to another. Threshold value associated with alarm time was set to detect change points as points that belong to the boundary B between two studied regions.

3. The results of brain tissues segmentation on T2-weighted MR perfusion images of a human head with abnormal brain anatomy using CUSUM filter for boundary pixels were evaluated. The evaluation results showed a good agreement with the reference standard with a Dice index values $0.9744 \pm 0.0216$, high values of sensitivity ($0.9765 \pm 0.0411$), specificity ($0.9911 \pm 0.0052$), and accuracy ($0.9881 \pm 0.0101$). Comparison with the state-of-the-art method showed reducing the error rate with respect to the false negative results in lesion regions. This gives the reasons to consider the proposed method as a very satisfactory for brain tissues segmentation and, at the same time, it is fully automated and does not require a large computation time. Therefore, the proposed method could significantly reduce the time and efforts required to obtain desirable results for perfusion ROI detection on T2-weighted MR perfusion images with abnormal brain anatomy.

## 7. REFERENCES

[1] G.H. Jahng, K.L.L. Ostergaard, F. Calamante, "Perfusion magnetic resonance imaging: a comprehensive update on principles and techniques," *Korean Journal of Radiology*, vol. 15, issue 5, pp. 554-577, 2014.

[2] B. Lanzman, and J.J. Heit, "Advanced MRI measures of cerebral perfusion and their clinical applications," *Topics in Magnetic Resonance Imaging*, vol. 26, issue 2, pp. 83-90, 2017.

[3] K. Welker, J. Boxerman, A. Kalnin, T. Kaufmann, M. Shiroishi, M. Wintermark, "ASFNR recommendations for clinical performance of MR dynamic susceptibility contrast perfusion imaging of the brain," *American Journal of Neuroradiology*, vol. 36, issue 6, pp. E41-E51, 2015.

[4] I. Galinovic, A.C. Ostwaldt, C. Soemmer, et. al., "Automated vs manual delineations of regions of interest – a comparison in commercially available perfusion MRI software," *BMC Medical Imaging*, vol. 12, no. 16, 2012, [Online]. Available from: www.biomedcentral.com/1471-2342/12/16.

[5] С.М. Алхімова, О.С. Железний, "Проблема автоматизації визначення зони уваги в перфузійних магнітно-резонансних дослідженнях," *сб. науч. тр. по материалам*

*международной науч.-практ. конф. "Сучасні напрямки теоретичних і прикладних досліджень '2015"*, Одесса, 2015, Вип. 1, Т. 4, с. 90-93.

[6] I. Galinovic, P. Brunecker, A.C. Ostwaldt, et. al., "Fully automated postprocessing carries a risk of substantial overestimation of perfusion deficits in acute stroke magnetic resonance imaging," *Cerebrovascular Diseases*, vol. 31, issue 4, pp. 408-413, 2011.

[7] S.M. Alkhimova, "Detection of perfusion ROI as a quality control in perfusion analysis," *Proceedings of the Conference Science, Research, Development. Technics and Technology,* Berlin, Germany, January 30, 2018, pp. 57-59.

[8] I. Despotovic, B. Goossens, W. Philips "MRI segmentation of the human brain: Challenges. methods, and applications," *Computational and Mathematical Methods in Medicine*, vol. 2015, article ID 450341, 23 pages, 2015.

[9] D. Selvaraj, R. Dhanasekaran, "MRI brain image segmentation techniques – A review," *Indian Journal of Computer Science and Engineering*, vol. 4, issue 5, pp. 364-381, 2013.

[10] M.A. Balafar, A.R. Ramli, M.I. Saripan, S. Mashohor, "Review of brain MRI image segmentation methods," *Artificial Intelligence Review*, vol. 33, issue 3, pp. 261-274, 2010.

[11] S. Tripathi, R.S. Anand, E. Fernandez, "A review of brain MR image segmentation techniques," in *Proceedings of the International Conference on Recent Innovations in Applied Science, Engineering & Technology (AET-2018)*, Mumbai, India, June 16-17, 2018, pp.62-69.

[12] С.М. Алхімова, М.М. Шарган, "Визначення оптимальних параметрів сегментації мозку на МРТ-зображеннях," *сб. науч. тр. по матер. науч.-практ. конф. "Современные проблемы и пути их решения в науке, транспорте, производстве и образовании '2014"*, Одесса, 2014, Вип. 2, Т. 7, с. 90-93.

[13] S. Datta, P.A. Narayana, "Automated brain extraction from T2-weighted magnetic resonance images," *Journal of Magnetic Resonance Imaging*, vol. 33, issue 4, pp. 822-829, 2011.

[14] S. Rajagopalan, R.A. Karwoski, R. Robb, "Robust fast automatic skull stripping of MRI-T2 data," *Proceedings of SPIE. Medical Imaging: Image Processing*, San Diego, CA, United States, February 13-17, 2005, vol. 5747, pp. 485-495.

[15] Z. Jin, A.L. Bertozzi, "Environmental boundary tracking and estimation using multiple autonomous vehicles," *Proceedings of the 2007 46th IEEE Conference on Decision and Control*, New Orleans, LA, USA, December 12-14, 2007, pp. 4918–4923.

[16] W. Liu, Y.E. Taima, M.B. Short, A.L. Bertozzi, "Multi-scale collaborative searching through swarming," *Proceedings of the 7th International Conference on Informatics in Control, Automation and Robotics (ICINCO)*, Funchal, Madeira, Portugal, June 15-18, 2010, pp. 222-231.

[17] A. Chen, T. Wittman, A.G. Tartakovsky, A.L. Bertozzi, "Image segmentation through efficient boundary sampling," *Proceedings of the 8th International Conference on Sampling Theory and Applications (SAMPTA'09)*, Marseille-Luminy, France, May 18-22, 2009, pp. Special-session.

[18] A. Chen, T. Wittman, A.G. Tartakovsky, A.L. Bertozzi, "Efficient boundary tracking through sampling," *Applied Mathematics Research eXpress*, vol. 2011, issue 2, pp. 182-214, 2011.

[19] A. Chen, "Improved boundary tracking by off-boundary detection," *Proceedings of SPIE. Remote Sensing: Image and Signal Processing for Remote Sensing XVIII*, Edinburgh, United Kingdom, September 24-27, 2012, vol. 8537, pp. 85370E1-7.

[20] E.S. Page, "A test for a change in a parameter occurring at an unknown point," *Biometrika*, vol. 42, issue 4, pp. 523-527, 1955.

[21] J.M. Lucas, and R.B. Crosier, "Fast initial response for CUSUM quality-control schemes: Give your CUSUM a head start," *Technometrics*, vol. 24, issue 3, pp. 199-205, 1982.

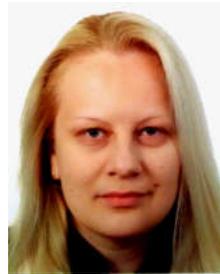

***Svitlana Alkhimova** is an associate professor at the Department of Biomedical Cybernetics, Igor Sikorsky Kyiv Polytechnic Institute, PhD in biological and medical devices and systems (2013). Areas of scientific interests are medical image processing and analysis, medical imaging.*

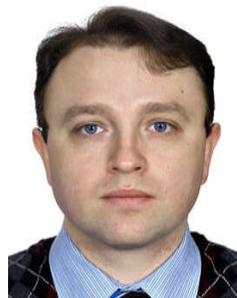

***Andrii Krenevych** is an associate professor at the Department of Mathematical Physics, Taras Shevchenko National University of Kyiv, PhD in differential equations (2008). Areas of scientific interests are stochastic differential equations, differential equations on time scales, image processing.*